\documentclass[10pt, conference]{IEEEtran}
\usepackage{hyperref}
\usepackage{graphicx}
\usepackage{amsthm,amssymb,amsmath}
\usepackage{bm}
\usepackage{gensymb}
\usepackage{multirow}
\usepackage[justification=centering]{caption}
\usepackage{color}
\usepackage{booktabs}
\usepackage{adjustbox}
\usepackage{breqn}
\usepackage{float}

\hypersetup{pdfborder=0 0 0}
\begin{document}

\title{Predicting Resilience with Neural Networks}

%Update author names here
\author{\IEEEauthorblockN{Karen da Mata$^1$, Priscila Silva$^1$ and Lance Fiondella$^1$}
\IEEEauthorblockA{$^1$Electrical and Computer Engineering, University of Massachusetts Dartmouth, MA, USA\\}
% \IEEEauthorblockA{$^2$Tandy School of Computer Science, University of Tulsa, OK, USA\\}
%\IEEEauthorblockA{$^2$Naval Air Systems Command, Patuxent River, MD, USA\\
Email: \{kalvesdamata, psilva4 and lfiondella\}@umassd.edu}
%}

\maketitle
\pagestyle{empty}

\begin{abstract}
\textcolor{black}{Resilience engineering studies the ability of a system to survive and recover from disruptive events, which finds applications in several domains. Most studies emphasize resilience metrics to quantify system performance, whereas recent studies propose statistical modeling approaches to project system recovery time after degradation. Moreover, past studies are either performed on data after recovering or limited to idealized trends. Therefore, this paper proposes three alternative neural network (NN) approaches including (i) Artificial Neural Networks, (ii) Recurrent Neural Networks, and (iii) Long-Short Term Memory (LSTM) to model and predict system performance, including negative and positive factors driving resilience to quantify the impact of disruptive events and restorative activities. Goodness-of-fit measures are computed to evaluate the models and compared with a classical statistical model, including mean squared error and adjusted R squared. Our results indicate that NN models outperformed the traditional model on all goodness-of-fit measures. More specifically, LSTMs achieved an over 60\% higher adjusted R squared, and decreased predictive error by 34-fold compared to the traditional method. These results suggest that NN models to predict resilience are both feasible and accurate and may find practical use in many important domains.}
\end{abstract}


\section{Introduction} 
\textcolor{black}{System resilience is the ability of a system or a process to survive and recover from disruptive events~\cite{hollnagel2006resilience, 2008Madni}. Early studies emphasize resilience metrics~\cite{hosseini2016review} to quantify system performance, while more recent studies~\cite{2022PriscilaDSN} propose resilience models to project system recovery time after failures. Past studies emphasizing resilience metrics or modeling are typically performed after recovering or limited to smooth trends. Real-world systems do not exhibit such simplified trends. Therefore, a general model capable of characterizing a broad cross-section of systems and processes to which resilience engineering is relevant would be beneficial.}

\textcolor{black}{Relevant research on quantitative resilience metrics includes Bruneau and Reinhorn~\cite{bruneau2007exploring} defined performance preserved relative to a baseline by measuring the area under the curve, while Yang and Frongopol~\cite{Yang2019} measured the performance lost due to a degrading stress as the area above the curve. Moreover, resilience models have been proposed with stochastic techniques such as Markov processes~\cite{2020DhulipalaFlint}, Bayesian networks~\cite{2022YinRen}, and Petri nets~\cite{2023YanDunnett}. Recently, Silva et al.~\cite{2022PriscilaDSN} proposed statistical models with multiple linear regression methods that include covariates, characterizing multiple deteriorations and recoveries associated with multiple shocks. Although these methods and models successfully quantify or characterize a system or a process performance under difficult conditions, they are either limited to idealized trends or single disruptive events, or contain various parameters to describe multiple shocks. Hence, machine learning techniques are a powerful alternative to these statistical models which can improve predictive accuracy and do not depend on idealized curve shapes.}

\textcolor{black}{In contrast to previous research, this paper considers three neural network (NN) models to predict system performance, including negative and positive factors driving deterioration and recovery in order to better understand the application domain and precisely track and predict the impact of disruptions and restorative activities, including artificial neural networks \cite{1943mcculloch} (ANN), (ii) recurrent neural networks \cite{1982hopfield} (RNN), and (iii) long-short term memory \cite{1997hoch} (LSTM). Multiple linear regression with interaction \cite{2022PriscilaDSN} (MLRI), a classical statistical method, is also applied to compare the predictive accuracy of the proposed models when tracking degradation and recovery, and predicting future changes in performance. All models are applied to historical data with $60\%$ and $70\%$ of data used for training. Our results suggest that LSTMs improve the adjusted R Squared over $60\%$ and reduce predictive error 34-fold compared to the statistical method, indicating a better model fit and a higher predictive accuracy.}

\textcolor{black}{The remainder of this paper is organized as follows: Section~\ref{sec:models} summarizes NN models to characterize resilience. Section~\ref{sec:ModelVal_FeatureSelection} describes the goodness-of-fit measures to validate the models and a feature selection technique considered to identify the most relevant subset of covariates. Section~\ref{sec:ill} provides illustrative examples and compares the model results. Section \ref{sec:conclusion} offers conclusions and future research.}

\section{Resilience Modeling}\label{sec:models}
\textcolor{black}{This section describes predictive resilience models, which have been developed to track and predict the performance of a system. Although the definition of the performance of a system is domain-dependent, it can be described as the level of accomplishment of a system or a task, which changes depending on disruptive events and restorative activities, also known as covariates, characterizing a resilience curve.}

\begin{figure}[ht!]
\centerline{\includegraphics[width=9.5cm,keepaspectratio]{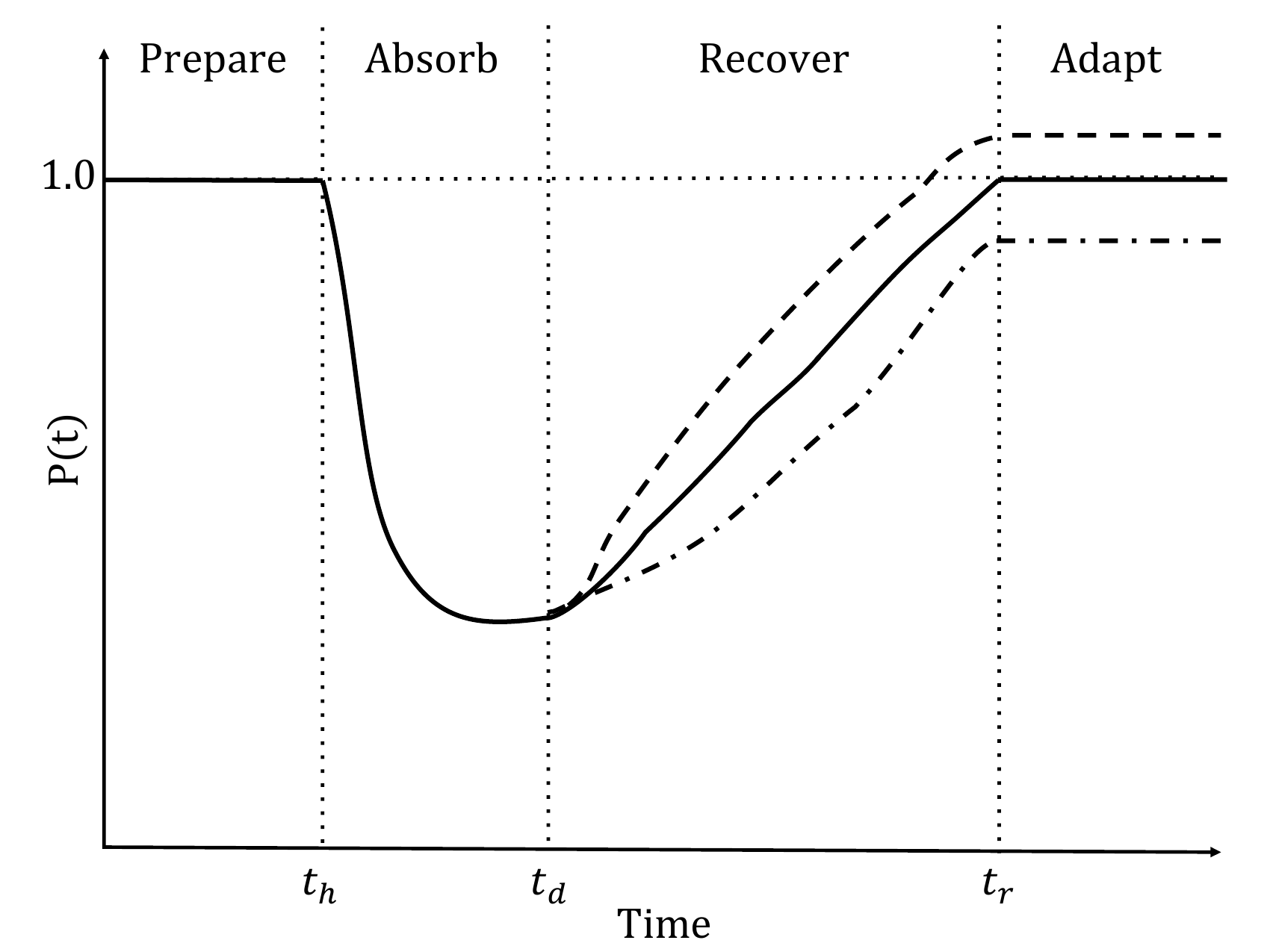}}
\caption{Canonical resilience curve.}\label{fig:resilience_curve}
\end{figure}

\textcolor{black}{Figure \ref{fig:resilience_curve} shows the four stages of a canonical resilience curve described by The National Academy of Sciences~\cite{2012TheNationalAcademy}, including (i) plan or prepare for, (ii) absorb, (iii) recover from, and (iv) adapt to actual or potential disruptive events. In the prepare stage, the system possesses a nominal performance $P(t)$ indicated by the dotted horizontal line until time $t_h$ when an initial disruptive event occurs. Then, the system transitions to the absorb stage, where the performance deteriorates until it reaches a minimum value at time $t_d$ and starts to improve in the recovery stage. Resilient systems recover smoothly to a new steady performance until time $t_r$ and enter into the adapt stage. Systems capable of adapting to their environment can recover to an improved (dashed) performance such as economic and computational systems. However, physical systems such as power generation may only exhibit recovery to nominal (solid) or degraded (dash-dotted) performance due to damaged equipment.}

\textcolor{black}{A discrete resilience curve incorporating covariates~\cite{2022PriscilaDSN} is}
\begin{equation}\label{eq:performance}
    P(i)=P(i-1) + \Delta P(i)
\end{equation}
\textcolor{black}{where $P(i)$ and $P(i-1)$ are the performance in the present and previous interval, and $\Delta P(i)$ is the change in performance predicted using the past~\cite{2022PriscilaDSN} and the proposed methods described in the following subsections.}

\subsection{Classical Statistical Modeling}
\textcolor{black}{Silva et al.~\cite{2022PriscilaDSN} characterized the change in performance with multiple linear regression with interaction as}
\begin{equation}
    \label{eq:changeinperformance_MLRI}
    \Delta P(i) = \beta_0 + \sum _{j=1}^m \beta_j X_j(i) + \sum _{j=1}^m \sum _{l=j+1}^m \beta_{j(m+l)} X_j(i)X_l(i)
\end{equation}
\textcolor{black}{where $\beta_0$ is the baseline change in performance, $X_1(i), X_2(i), \dots, X_m(i)$ the $m$ covariates documenting the magnitude of degrading shocks or amount of efforts dedicated to restore performance, $\beta_1, \beta_2, \dots, \beta_m $ the coefficients characterizing the impact of the covariates, and $\beta_{j(m+l)}$ the interaction between two covariates.}

\subsection{Neural Networks}
\textcolor{black}{A neural network~\cite{1997smith} is a machine-learning approach commonly used in solving pattern recognition and non-linear problems, since it does not impose assumptions about an underlying probability distribution or functional form, making it widely applicable to numerous real-world applications.}

The architecture of a neural network model consists of nodes, also known as neurons, an input layer, an output layer, and one or more hidden layers. \textcolor{black}{In each hidden layer, a predefined number of neurons process the information from the previous layer. An activation function transforms this information, which is transferred to the next layer. In most cases, the neurons are fully connected between layers. Each connection possesses a weight, while each hidden layer and the output layer also possess a bias. The network weights and biases are initialized randomly and estimated using historical data during training with an optimization algorithm. In this learning process, the network processes the entire training data set several times, known as epochs. During each epoch, the optimizer adjusts the weights and biases after each observation is processed to minimize the error between actual data and the output of the network. Once the training is complete, the model can be used to predict by providing new input.}

\begin{figure}[!ht]
\centerline{\includegraphics[width=9.5cm,keepaspectratio]{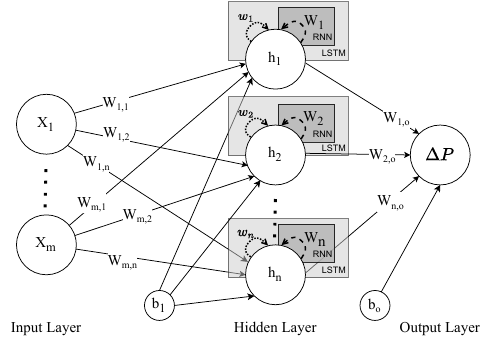}}
\caption{Topology of the neural networks.}\label{fig:nn-archicture}
\end{figure}

 \textcolor{black}{Figure \ref{fig:nn-archicture} shows a combination of three topology of the NN models considered in this paper. The ANN architecture is represented by the network disregarding the two loops in the hidden layer. The RNN architecture includes the ANN architecture as well as the dashed loop in the hidden layer. Thus, the LSTM architecture consists of the entire network considering both dashed and dotted loops in the hidden layer.}

\noindent \textcolor{black} {The input layer in Figure \ref{fig:nn-archicture} consists of $m$ neurons, one for each covariate $X_1(i),X_2(i), \dots, X_m(i)$ at time step $i$, which are interconnected with the $n$ neurons in the hidden layer, and $W_{1,1},W_{1,2}, \dots, W_{m,n}$ are the weights associated with the connection between the neurons of the input and the hidden layer. The hidden states $h_1(i), h_2(i), \dots, h_n(i)$ at time step $i$, are defined by the NN model applied, which are introduced in the following subsections. $b_1$ is the bias of the hidden layer, $W_1, W_2, \dots, W_n$ the weights associated with dashed recurrent loops included in the RNN and LSTM models, and $w_1,w_2, \dots, w_n$ the weights associated with the cell states included only in the LSTM model.} 

The output layer of the model consists of a single neuron characterizing the change in performance at time step $i$ as
% , \textcolor{black}{characterized as the weighted summation of the hidden layer outputs at time step $i$ plus the bias $b_o$ of the output layer such that}
\begin{equation}\label{eq:changeinperformance_NN}
    \Delta P(i) = \sum _{k=1}^n W_{k,o} h_k(i) + b_o
\end{equation}
where $W_{k,o}$ is the weight of the connection between the $k^{th}$ neuron of the hidden layer and the output layer, and $b_o$ the bias of the output layer.

\subsubsection{Artificial Neural Network (ANN)}
The artificial neural network~\cite{1943mcculloch} is the simplest case of a NN, \textcolor{black}{where each neuron in the hidden layer receives as input the summation of the weighted input neurons and a bias $b_1$ of the hidden layer. An activation function $\alpha$ transforms this summation introducing non-linearity to the network. Then, the output $h_k(i)$ of the $k^{th}$ neuron in the hidden layer at time step $i$ is} 
\begin{equation}
\label{eq:ann}
    h_k (i)=\alpha\left(\sum_{j=1}^m W_{j,k} X_j(i)+b_1\right)
\end{equation}
\textcolor{black}{where $W_{j,k}$ is the weight associated with the $j^{th}$ node of the input layer and the $k^{th}$ node in the hidden layer, $X_j(i)$ is the $j^{th}$ covariate in the present time step.}

\textcolor{black}{To avoid vanishing or exploding gradient problems due to the activation function, the single hidden layer of this model uses the ReLU activation function}
\begin{equation}
\label{eq:relu}
    \text{ReLU} (x) = \begin{cases}
                          x & \text{if } x>0 \\
                          0 & \text{Otherwise}
                    \end{cases}
\end{equation}
\textcolor{black}{and the input and output layers do not require activation functions. Thus, the change in performance is modeled with ANNs by substituting Equation (\ref{eq:ann}) and (\ref{eq:relu}) into Equation (\ref{eq:changeinperformance_NN}).}

\subsubsection{Recurrent Neural Network (RNN)} 

\textcolor{black}{A recurrent neural network \cite{1982hopfield} is an extension of the ANN that includes both the current time step as well as the previous output to make predictions. The dashed loop pointing back to each neuron in the hidden layer in Figure \ref{fig:nn-archicture} indicates the previous output being passed into each neuron, also known as the previous hidden state $h_k(i-1)$ of the $k^{th}$ neuron in the hidden layer. Thus, the hidden state $h_k(i)$ in the current time step is}
\begin{equation}\label{eq:hk_rnn}
    h_k(i) = \alpha \left (\sum_{j=1}^m W_{j,k} X_j(i) + W_k h_{k}(i-1)+b_1\right)
\end{equation}
\textcolor{black}{where $W_k$ is the weight associated with the recurrent portion of the $k^{th}$ node of the hidden layer. The hidden layer also uses the ReLU activation function. Thus, the change in performance is modeled with RNNs by replacing Equation (\ref{eq:hk_rnn}) and (\ref{eq:relu}) into Equation (\ref{eq:changeinperformance_NN}).}

\subsubsection{Long Short-Term Memory (LSTM)}

\textcolor{black}{Long short-term memory \cite{1997hoch} modifies the ANN/RNN architecture to address the vanishing and exploding gradient problems in which the gradients of the network tend to zero or infinity during training, by introducing new features enforcing constant error flow throughout the layers. Besides taking the input and hidden states as RNNs, LSTM introduces a ``cell state'' ($c_k$) which holds important information during an arbitrary time interval.}

\begin{figure}[!ht]
\centerline{\includegraphics[width=9.5cm,keepaspectratio]{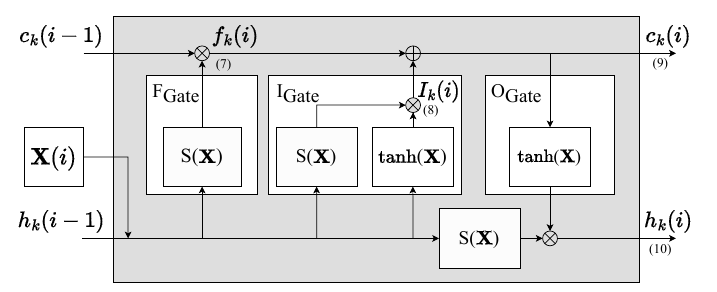}}
\caption{Example of a $k^{th}$ LSTM cell.}\label{fig:lstm-cell}
\end{figure}

\textcolor{black}{Figure \ref{fig:lstm-cell} shows an example of a LSTM cell containing memory cells and three gates, including the forget ($F_{Gate}$), input ($I_{Gate}$), and output ($O_{Gate}$) gates.}

\noindent\textcolor{black}{The $\oplus$ operator in Figure \ref{fig:lstm-cell} represents element wise addition and $\otimes$ vector multiplication, $c_k(i-1)$ the previous $k^{th}$ cell value, and $\textbf{X}(i)$ the current input vector. The LSTM model uses two activation functions, the hyperbolic tangent $\tanh(x)$ and the logistic $S(x)$. The $\tanh(x)$ processes an input and outputs a number between $-1$ and $1$, while the $S(x)$ outputs a number between $0$ and $1$, indicating if information should be ignored $(0)$ or kept $(1)$.}

\textcolor{black}{On the left side of Figure \ref{fig:lstm-cell}, the $F_{Gate}$ uses the logistic activation function to decide if the state of the previous cell should be remembered or ignored. Thus, the forget gate output $f_k(i)$ of the $k^{th}$ neuron in the hidden layer is}
\begin{equation} 
    f_k (i) = c_k(i-1)\times S\left(\sum_{j=1}^m W_{j,k} X_j(i) + W_k h_{k}(i-1)+b_1 \right)
\end{equation}

\textcolor{black}{The $I_{Gate}$ shown in the middle of Figure \ref{fig:lstm-cell} multiplies the output of both activation functions to decide which information should be stored in the cell state $c_k(i)$ resulting in}
\begin{dmath}
  I_k(i) = S\left(\sum_{j=1}^m W_{j,k} X_j(i) + W_k h_{k}(i-1)+b_1 \right) \times \tanh\left(\sum_{j=1}^m W_{j,k} X_j(i) + W_k h_{k}(i-1)+b_1 \right)
\end{dmath}

\textcolor{black}{Then, the updated cell state $c_k(i)$ is}
\begin{equation}
    c_k(i) = f_k(i) + I_k(i)
\end{equation}

\textcolor{black}{The final gate $O_{Gate}$ on the right side of Figure \ref{fig:lstm-cell} computes the $k^{th}$ hidden state $h_k(i)$ at the current time step as}
\begin{dmath}\label{eq:hk_lstm}
  h_k(i) = S\left(\sum_{j=1}^m W_{j,k} X_j(i) + W_k h_{k}(i-1)+b_1 \right) \times \tanh\left(w_k \times c_k(i) + b_1\right)
\end{dmath}
\textcolor{black}{where $w_k$ is a weight associated with the cell state, and $c_k(i)$ and $h_k(i)$ are the final outputs of the $k^{th}$ LSTM cell in the hidden layer. Thus, the change in performance is modeled with LSTMs by replacing Equation (\ref{eq:hk_lstm}) into Equation (\ref{eq:changeinperformance_NN}).}

\section{Model Validation and Feature Selection}
\label{sec:ModelVal_FeatureSelection}
\textcolor{black}{This section describes measures to assess the performance of the models described in Section \ref{sec:models} and a feature selection technique to identify the most relevant subset of covariates.}

\subsection{Goodness-of-fit Measures}
\label{sec:modelassessment}

\textcolor{black}{Goodness-of-fit measures assess how well the model performs on a given data set. Once the models described in Section \ref{sec:models} are fitted/trained and the predictions of the change in performance ($\Delta P$) are made, the results of Equation (\ref{eq:changeinperformance_MLRI}) and (\ref{eq:changeinperformance_NN})  can be replaced in Equation (\ref{eq:performance}) to estimate the system performance ($\hat{P}$) to evaluate the models.}

\textcolor{black}{Statistical and machine learning models require different data splits. Multiple linear regression with interaction divides the data set into two parts, training and testing. While machine learning models split the data into three parts, training, validation and testing. Thus, all models are fitted/trained with the training data set of $n-l$ points. The regression model uses the remainder $l$ points for testing, while the NN models divide the remaining $l$ points into two equal parts since they require a validation data set to evaluate the models during training to select the best hyperparameters, such as number of neurons and layers. Then, the testing data set of $\frac{l}{2}$ points provide an unbiased evaluation of the NN model selected after training.}

\textit{Predictive Mean Squared Error (PMSE)}
\textcolor{black}{~\cite{2013kleinbaum} computes the mean discrepancy of the model estimate from the actual data considering the testing data set.}
\begin{equation}
    \text{PMSE}=\frac{1}{l_{m}} \sum _{i=n-l_{m}+1}^n \left(\hat{P}(i)-P(i)\right)^2
\end{equation}
\textcolor{black}{where $l_{m}=l$ for regression, and $\frac{l}{2}$ for the NNs, and $\hat{P}(i)$ and $P(i)$ are the predicted and expected performance at time $i$. Two special cases are the \textit{Validation Mean Squared Error (VMSE)} where the error considers only the validation data set presented in the NN models, and the \textit{Mean Squared Error (MSE)} considers the entire data set for both models.}

\textit{Mean Absolute Percentage Error (MAPE)}
\textcolor{black}{\cite{2016Myttenaere} measures the mean accuracy of time-dependent problems.}
\begin{equation}
    MAPE = \frac{100}{n} \sum _{i=1}^n \bigg\lvert\frac{P(i)-\hat{P}(i)}{P(i)}\bigg \rvert
\end{equation}

\textcolor{black}{For all error measures (PMSE, VMSE, MSE and MAPE), smaller values are preferred since they indicate a better model fit compared with other models.}

\textit{Adjusted Coefficient of Determination ($r^2_{adj}$)}
\textcolor{black}{ \cite{1995Anil} measures the variation in the dependent variable that is explained by $m$ independent variables incorporated into the model, quantifying the degree of linear correlation between the empirical performance and the model predictions.} 
\begin{equation}
    r^2_{adj}=1-\left(1-\frac{\text{SSY}-\text{SSE}}{\text{SSY}}\right)\left(\frac{n-1}{n-m-1}\right)
\end{equation}
where 
\begin{dmath}
  \text{SSY} = \sum _{i=1}^n \left(P(i) - \overline{P}(i)\right)^2
\end{dmath}
\textcolor{black}{is the sum of squares error associated with the naive predictor
$\overline{P}(i)$, and}
\begin{equation}
    \text{SSE} = \sum _{i=1}^n \left(\hat{P}(i)-P(i)\right)^2
\end{equation}
\textcolor{black}{is the sum of squares of the residual between the predicted $\hat{P}(i)$ and expected performance $P(i)$. A value of $r^2_{adj}$ closer to 1.0 indicates a strong relationship between the data and the model. Negative or low $r_{adj}^2$ values indicate no or weak linear relationship, which may be due to poor predictions or model fit that result in a large SSE.}

\subsection{A Hybrid Feature Selection Technique}
\label{sec:featureSelection}

\textcolor{black}{Hybrid feature selection techniques \cite{2010Min} are widely applied in machine learning problems to reduce the number of model inputs and decrease the possibility of under or overfitting. There are two steps to select the most relevant subset of covariates. The first step performs a forward selection search ranking possible subsets of covariates using a heuristic ``merit'' function. The second step trains models using the highest-ranked subsets from the first step and evaluates the models according to the goodness-of-fit measures.}

\textcolor{black}{The first step applies a correlation-based feature selection (CFS) technique \cite{1999Hall} to identify a subset of relevant covariates highly correlated with the expected output ($\Delta P$) but uncorrelated with the other covariates to avoid including redundant attributes. The heuristic ``merit'' evaluation function}
\begin{equation}
\label{eq:merit}
    M_s = \frac{k \overline{r_{co}}}{\sqrt{k+k(k-1)\overline{r_{cc}}}}
\end{equation}
\textcolor{black}{ranks a subset $S$ of $k$ covariates, where $\overline{r_{co}}$ is the mean of the correlation between the covariates in the subset $S$ and the expected output $\Delta P$, and $\overline{r_{cc}}$ the mean of the inter-correlation between the covariates in the subset $S$.}

\textcolor{black}{The forward selection search starts by evaluating one covariate at a time using Equation (\ref{eq:merit}). Then, the search evaluates the highest-ranked covariate with each remaining covariate, and chooses the subset with the highest score. This process stops when the change in the merit score decreases by more than 0.01 or reaches the maximum number of covariates. The second step of the hybrid feature selection creates and trains models using the highest-ranked subsets resulting from the CFS algorithm as input, and evaluates these models according to the goodness-of-fit measures introduced in Section \ref{sec:modelassessment}. The subset of covariates that achieves the highest $r^2_{adj}$ and the smallest overall error is then chosen as the best subset of covariates for the application.}

\section{Illustrations}\label{sec:ill}

\textcolor{black}{This section illustrates the application of machine learning models described in Section \ref{sec:models}, including artificial neural network, recurrent neural network, and long short-term memory to predict the systems change in performance. Predictions made by the NN models are compared with the traditional statistical multiple linear regression with interaction model results to evaluate the effectiveness of the approaches through the goodness-of-fit measures described in Section \ref{sec:modelassessment}.}

\textcolor{black}{The proposed resilience models are illustrated using the most recent recession in the U.S. that began in March, 2020 due to the COVID-19 pandemic. In this case, performance is the normalized number of adults eligible to work in the United States, where time step zero corresponds to peak employment prior to a period of job loss and recovery. In order to identify activities that have a high impact on job losses or recovery, twenty-one covariates regarding statistical information and relevant factors to the COVID-19 pandemic were collected from January 2020 to November 2022, and normalized by dividing the values of each covariate by the maximum value observed for that covariate. To promote reproducibility, these covariates are available in a public GitHub repository~\cite{Website-GH-Repository}.}

\textcolor{black}{For each NN model considered, the number of neurons in the hidden layer was varied between 1 to 15 neurons, due to the limited size of the data set. The models were trained with the Adam optimizer~\cite{kingma2014adam} and an upper limit of 1000 epochs with an earlier stopping condition when the change in the loss did not improve by more than 0.0001 for ten epochs. Three possible values for the learning rate were also tested, namely $\alpha$ = \{$10^{-2}$, $10^{-3}$, $10^{-4}$\}. Ultimately, an $\alpha=0.01$ was selected to achieve a good fit while avoiding overfitting. For each combination, the model was trained and tested 50 times, and the average of the results was compared through to the measures described in Section \ref{sec:modelassessment}. The data set was split into three parts, training-validation-testing, and two splits were considered, 60-20-20 and 70-15-15. In order to conduct an objective comparison, the MLRI models were fitted using maximum likelihood estimation~\cite{hogg1995introduction} with 60\% and 70\% of the data and the remainder was used for testing.}

\textcolor{black}{Table \ref{tbl:feature-subset} shows the result of the first part of the hybrid feature selection technique described in Section \ref{sec:featureSelection} applied to the 2020 U.S. recession data set. The order in which the covariates were selected is X$_{19}$ (\textit{Industrial Production}), X$_{14}$ (\textit{Workplace Closures}), X$_{4}$ (\textit{New Orders Index}), X$_{7}$ (\textit{Unemployment Benefits}), and X$_{6}$ (\textit{Consumer Activity}).}
\begin{table}[ht!]
\setcounter{table}{0}
    \caption{Ranking of covariates subset using CFS algorithm.}
    \label{tbl:feature-subset}
    \centering
    \begin{tabular}{@{}lll@{}}
    \toprule
    \multicolumn{1}{c}{Covariates Subset} & \multicolumn{1}{c}{\(k\)} & \multicolumn{1}{c}{\(M_s\)} \\ \midrule
    X$_{19}$ & \multicolumn{1}{c}{1} & \multicolumn{1}{r}{0.5567882} \\
    X$_{19}$, X$_{14}$ & \multicolumn{1}{c}{2} & \multicolumn{1}{r}{0.6151150} \\
    X$_{19}$, X$_{14}$, X$_{4}$ & \multicolumn{1}{c}{3} & \multicolumn{1}{r}{0.6257571} \\
    X$_{19}$, X$_{14}$, X$_{4}$, X$_{7}$ & \multicolumn{1}{c}{4} & \multicolumn{1}{r}{0.6208308} \\
    X$_{19}$, X$_{14}$, X$_{4}$, X$_{7}$, X$_{6}$ & \multicolumn{1}{c}{5} & \multicolumn{1}{r}{0.5959661} \\ \bottomrule
    \end{tabular}
\end{table}

\noindent\textcolor{black}{As shown in Table \ref{tbl:feature-subset}, the subset of covariates that has the highest merit score included three covariates, X$_{19}$, X$_{14}$, X$_{4}$, and when a fourth covariate was included, the merit score started to decrease. The subset with more than four covariates decreased by more than $0.01$, and was hence disregarded.}

\textcolor{black}{Table~\ref{tbl:covid_results} reports the goodness-of-fit comparison and the architecture of the best combinations of the models discussed in Section \ref{sec:models} for the four subsets selected in the first part of the hybrid feature selection method. Most models exhibited inconsistent results when comparing both data splits, demonstrating that these models perform differently depending on the training data size, which is not always expected. For these reasons, the architecture of each model that produced the most similar results in both data splits was chosen, shown in bold.} 

\begin{table*}[h]
\setcounter{table}{1}
    \caption{Architecture and goodness-of-fit comparison on U.S. recession data set. (Best results of each model in bold)}
    \centering
    \begin{adjustbox}{scale=0.85}
    \begin{tabular}{@{}clcrrrrrrllrrrr@{}}
    \toprule
    Model & \multicolumn{1}{c}{\begin{tabular}[c]{@{}c@{}}Covariates \\ Subset\end{tabular}} & Neurons & \multicolumn{2}{c}{PMSE} & \multicolumn{2}{c}{VMSE} & \multicolumn{2}{c}{MSE} & \multicolumn{2}{c}{MAPE} & \multicolumn{2}{c}{\begin{tabular}[c]{@{}c@{}}Adjusted \\ R Squared\end{tabular}} & \multicolumn{2}{c}{\begin{tabular}[c]{@{}c@{}}Average \\ Epochs\end{tabular}} \\ \midrule
    \multicolumn{1}{l}{Training Split} &  & \multicolumn{1}{l}{} & \multicolumn{1}{c}{60} & \multicolumn{1}{c}{70} & \multicolumn{1}{c}{60} & \multicolumn{1}{c}{70} & \multicolumn{1}{c}{60} & \multicolumn{1}{c}{70} & \multicolumn{1}{c}{60} & \multicolumn{1}{c}{70} & \multicolumn{1}{c}{60} & \multicolumn{1}{c}{70} & \multicolumn{1}{c}{60} & \multicolumn{1}{c}{70} \\\midrule
    \multirow{4}{*}{MLRI} & X$_{19}$ & - & 0.0018884 & 0.0023357 & \multicolumn{1}{c}{-} & \multicolumn{1}{c}{-} & 0.001042 & 0.000950 & 2.40 & 2.28 & 0.2650 & 0.3298 & \multicolumn{1}{c}{-} & \multicolumn{1}{c}{-} \\
     & \textbf{X$_{19}$, X$_{14}$} & \textbf{-} & \textbf{0.0002884} & \textbf{0.0004306} & \multicolumn{1}{c}{\textbf{-}} & \multicolumn{1}{c}{\textbf{-}} & \textbf{0.000538} & \textbf{0.000553} & \textbf{1.70} & \textbf{1.73} & \textbf{0.6087} & \textbf{0.5978} & \multicolumn{1}{c}{\textbf{-}} & \multicolumn{1}{c}{\textbf{-}} \\
     & X$_{19}$, X$_{14}$, X$_{4}$ & - & 0.0138236 & 0.0048261 & \multicolumn{1}{c}{-} & \multicolumn{1}{c}{-} & 0.006013 & 0.001721 & 5.31 & 2.85 & -3.5224 & -0.2943 & \multicolumn{1}{c}{-} & \multicolumn{1}{c}{-} \\
     & X$_{19}$, X$_{14}$, X$_{4}$, X$_{7}$ & - & 0.0022365 & 0.0000272 & \multicolumn{1}{c}{-} & \multicolumn{1}{c}{-} & 0.000941 & 0.000044 & 2.01 & 0.54 & 0.2679 & 0.9657 & \multicolumn{1}{c}{-} & \multicolumn{1}{c}{-} \\\midrule
    \multirow{4}{*}{ANN} & X$_{19}$ & 3 & 0.0000481 & 0.0007419 & 0.0000838 & 0.0000604 & 0.000914 & 0.001090 & 2.19 & 2.66 & 0.3554 & 0.2312 & 423 & 450 \\
     & X$_{19}$, X$_{14}$ & 5 & 0.0000099 & 0.0010889 & 0.0000087 & 0.0000887 & 0.000333 & 0.000480 & 1.02 & 1.63 & 0.7575 & 0.6503 & 429 & 662 \\
     & \textbf{X$_{19}$, X$_{14}$, X$_{4}$} & \textbf{3} & \textbf{0.0000247} & \textbf{0.0000731} & \textbf{0.0000085} & \textbf{0.0000094} & \textbf{0.000245} & \textbf{0.000272} & \textbf{0.81} & \textbf{0.81} & \textbf{0.8154} & \textbf{0.7952} & \textbf{546} & \textbf{609} \\
     & X$_{19}$, X$_{14}$, X$_{4}$, X$_{7}$ & 14 & 0.0000046 & 0.0009769 & 0.0000043 & 0.0002925 & 0.000321 & 0.000216 & 0.96 & 1.04 & 0.7506 & 0.8320 & 855 & 870 \\ \midrule
    \multirow{4}{*}{RNN} & X$_{19}$ & 5 & 0.0003018 & 0.0014525 & 0.0003708 & 0.0001064 & 0.000652 & 0.000953 & 1.88 & 2.39 & 0.5400 & 0.3281 & 388 & 421 \\
     & X$_{19}$, X$_{14}$ & 7 & 0.0000562 & 0.0001318 & 0.0000352 & 0.0002553 & 0.000322 & 0.000400 & 1.15 & 1.45 & 0.7656 & 0.7086 & 388 & 471 \\
     & X$_{19}$, X$_{14}$, X$_{4}$ & 13 & 0.0000296 & 0.0001430 & 0.0000697 & 0.0000260 & 0.000400 & 0.000323 & 1.19 & 1.19 & 0.6990 & 0.7568 & 407 & 400 \\
     & \textbf{X$_{19}$, X$_{14}$, X$_{4}$, X$_{7}$} & \textbf{12} & \textbf{0.0000498} & \textbf{0.0001766} & \textbf{0.0000415} & \textbf{0.0000182} & \textbf{0.000355} & \textbf{0.000320} & \textbf{1.21} & \textbf{1.05} & \textbf{0.7241} & \textbf{0.7512} & \textbf{406} & \textbf{457} \\ \midrule
    \multirow{4}{*}{LSTM} & X$_{19}$ & 1 & 0.0000653 & 0.0004066 & 0.0001530 & 0.0000261 & 0.000307 & 0.000856 & 1.19 & 2.24 & 0.7838 & 0.3967 & 275 & 283 \\
     & X$_{19}$, X$_{14}$ & 2 & 0.0003256 & 0.0000072 & 0.0001011 & 0.0000094 & 0.000477 & 0.000252 & 1.66 & 0.90 & 0.6532 & 0.8169 & 415 & 331 \\
     & X$_{19}$, X$_{14}$, X$_{4}$ & 1 & 0.0009285 & 0.0006370 & 0.0001387 & 0.0000553 & 0.000426 & 0.000407 & 1.67 & 1.67 & 0.6799 & 0.6940 & 514 & 490 \\
     & \textbf{X$_{19}$, X$_{14}$, X$_{4}$, X$_{7}$} & \textbf{7} & \textbf{0.0000205} & \textbf{0.0000006} & \textbf{0.0000035} & \textbf{0.0000031} & \textbf{0.000015} & \textbf{0.000017} & \textbf{0.32} & \textbf{0.29} & \textbf{0.9885} & \textbf{0.9867} & \textbf{641} & \textbf{598} \\ \bottomrule
    \end{tabular}
    \end{adjustbox}
    \label{tbl:covid_results}
\end{table*}

\textcolor{black}{Table~\ref{tbl:covid_results} shows that the best MLRI model with two covariates achieved a $r^{2}_{\text{adj}}$ of $0.6032$ on average, as well as a low PMSE, MSE and a MAPE of $1.72$ on average. The best ANN model with three covariates and three neurons in the hidden layer achieved a $(\frac{0.8053}{0.6032}) = 33.5\%$ higher $r^{2}_{\text{adj}}$ and a $(\frac{1.72}{0.81}) = 2.12$ times lower MAPE, compared to the best MLRI. The best RNN model had four covariates and twelve neurons, achieving a $(\frac{0.7376}{0.6032}) = 22.2\%$ higher $r^{2}_{\text{adj}}$ than the MLRI, but the highest VMSE compared to the other NNs. The best LSTM with four covariates and seven neurons exhibited an $(\frac{0.9876}{0.6032}) = 63.72\%$ improvement on the $r^{2}_{\text{adj}}$, and a $(\frac{(2.884\cdot 10^{-4}+4.306\cdot 10^{-4})/2}{(2.048\cdot 10^{-5}+0.057\cdot 10^{-5})/2}) = 34.07$ times smaller PMSE, when compared to the best MLRI. The improvements made by LSTM justify the additional epochs on required.}

\begin{figure}[!ht]
\centerline{\includegraphics[width=9.5cm,keepaspectratio]{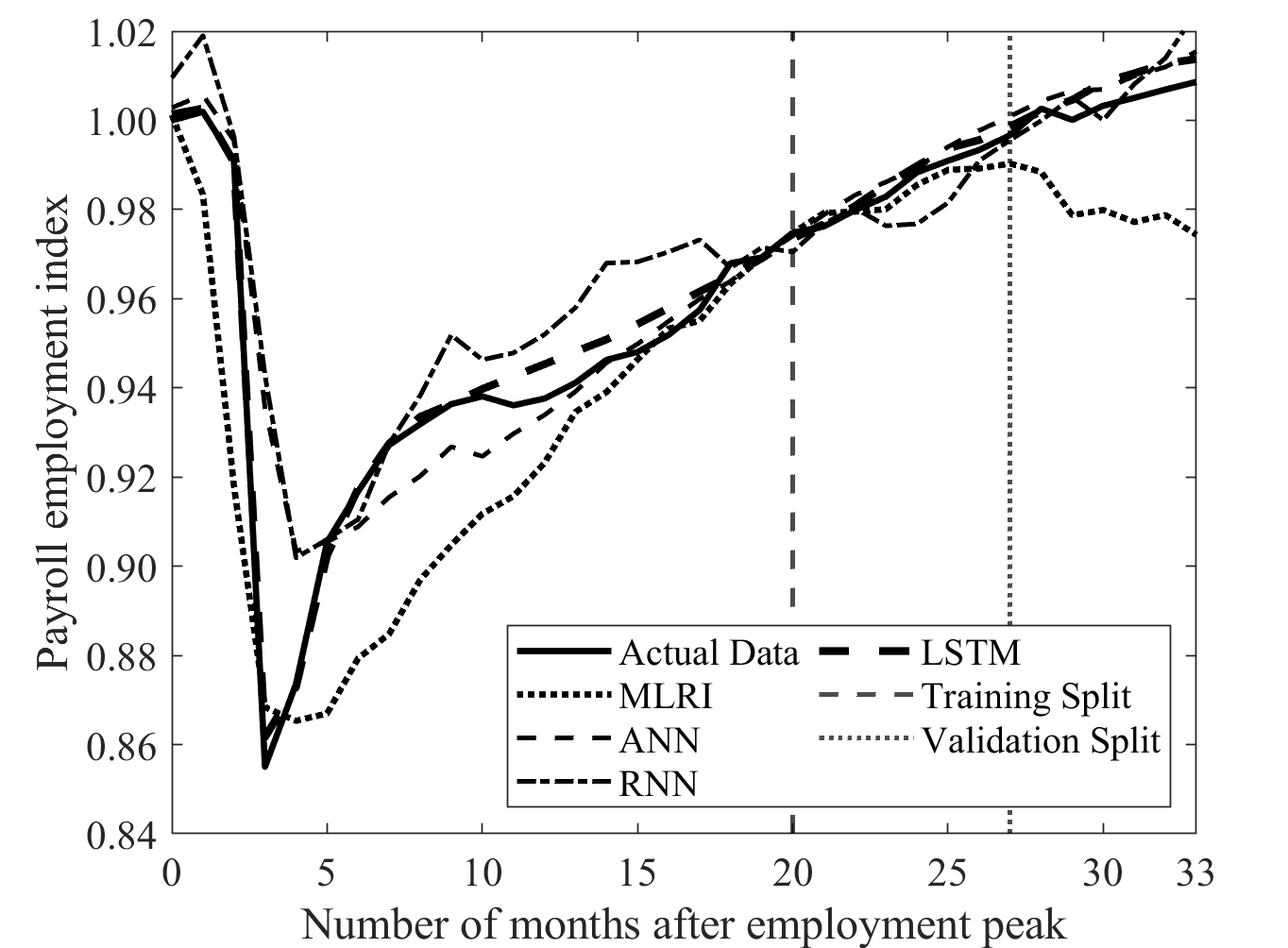}}
    \caption{Model fit of the best models using 60\% of the data for training.}\label{fig:modelfits}
\end{figure}

\textcolor{black}{Figure \ref{fig:modelfits} shows the normalized number of individuals employed (solid) and the model of best fit. The two vertical lines represent the time step in which the training and validation splits end. Since both considered splits have shown similar results for the chosen models, only the model fit using 60\% of the data for training is shown for clarity.}

\noindent \textcolor{black}{In Figure \ref{fig:modelfits}, the MLRI model underestimated the data in most phases, which may be due to the amount of data used to estimate the parameters, since mathematical approaches usually use 80\% or 90\% of the data for parameters estimation. The ANN model performed exceptionally well in the validation and testing phases and followed the data relatively well at the beginning of the training phase. However, the model did not track the degradation and underestimated the recovery. On the other hand, the RNN model oscillated between overestimating and underestimating in every phase. As explained in Section \ref{sec:models}, ANNs and RNNs are prone to gradient problems resulting in poor performance in some time-dependent problems. Meanwhile, the LSTM model characterized the data well in each phase. Thus, the LSTM's more complex architecture enables it to accommodate fluctuations better.}

\section{Conclusion and Future Research}\label{sec:conclusion}

\textcolor{black}{This paper presented three alternative neural network approaches, including (i) Artificial Neural Network, (ii) Recurrent Neural Network, and (iii) Long Short-Term Memory, to model and predict system resilience considering disruptive events and restorative activities that characterize the degradation and the recovery in system performance. The neural network models and a traditional model (Multiple Linear Regression with interaction) were applied to a historical data set with $60\%$ and $70\%$ of the data used for training. The results indicated that neural network approaches outperformed the multiple linear regression with interaction model in every stage of the analysis, including tracking degradation and recovery and predicting future changes in performance. Specifically, LSTMs exhibited an improvement of over $60\%$ in the adjusted R squared and a 34-fold reduction in predictive error.} 

\textcolor{black}{Future research will explore more challenging data sets including multiple shocks and alternative neural network models such as gated recurrent units.}

\section*{Acknowledgment}
This paper was published and presented at the $28^{th}$ ISSAT International Conference on Reliability \& Quality in Design in August 2023, which will be indexed in Elsevier Scopus.

\bibliographystyle{IEEEtran}
\bibliography{bibREU}

\end{document}